\documentclass[12pt,manuscript,authoryear]{aastex}

\usepackage{graphicx}

\begin{document}

\title{Giant Planet Formation by Disk Instability: 
A Comparison Simulation With An Improved Radiative Scheme}
  
\author{{\it Short Title: Disk Instability Comparison ~~~~Article Type: Journal}}

\author{Kai Cai}
\affil{Department of Chemistry \& Physics, Purdue University Calumet,  Hammond, IN 46323}
\email{Kai.Cai.6@nd.edu}

\and

\author{Megan K. Pickett}
\affil{Department of Physics, Lawrence University, Appleton, WI 54912}
\email{megan.pickett@lawrence.edu}

\and 

\author{Richard H. Durisen}
\affil{Astronomy Department, Indiana University,
    Bloomington, IN 47405}
\email{durisen@astro.indiana.edu}

\and

\author{Anne M. Milne}
\affil{Department of Physics, Lawrence University, Appleton, WI 54912}
\email{mlinea@lawrence.edu}

\begin{abstract}

There has been disagreement about whether cooling in protoplanetary disks can be sufficiently fast to induce the formation of gas giant protoplanets via gravitational instabilities. Simulations by our own group and others indicate that this method of planet formation does not work for disks around young, low-mass stars inside several tens of AU, while simulations by other groups show fragmentation into protoplanetary clumps in this region. To allow direct comparison in hopes of isolating the cause of the differences, we here present a high resolution three-dimensional hydrodynamics simulation of a protoplanetary disk, where the disk model, initial perturbation, and simulation conditions are essentially identical to those used in a recent set of simulations by \citet[][hereafter B07]{boss07}. As in earlier papers by the same author, B07 purports to show that cooling is fast enough to produce protoplanetary clumps. Here, we evolve the same B07 disk using an improved version of one of our own radiative schemes and find that the disk does not fragment in our code but instead quickly settles into a state with only low amplitude nonaxisymmetric structure, which persists for at least several outer disk rotations. We see no rapid radiative or convective cooling. We conclude that the differences in results are due to different treatments of regions at and above the disk photosphere, and we explain at least one way in which the scheme in B07 may lead to artificially fast cooling.

\end{abstract}


\label{firstpage}

\section{Introduction}

The issue of Jupiter and Saturn's provenance, as well as that of the known  extrasolar gas giant planets, remains a subject of intense study.  Over the last decade, several computational groups,  using an array of different techniques and disk models, have  examined the {\sl disk instability} mechanism, where gas-phase gravitational instabilities (GI's) produce spiral waves that fragment into protoplanetary clumps \citep{boss97,boss98, boss00, boss01, boss02, boss03, boss04, boss05, boss07, boss09, pickett98, pickett00a, pickett00b, pickett03, nelson98, nelson00, nelson06, gammie01, mayer02, mayer04, mayer07, johnson03, rice03, rice05, mejia05, cai06, cai08, boley06, boley07, stam08, forg09}.   Researchers agree that GI's are triggered when disks are massive and cold and that, once GI's occur, cooling on time scales comparable to the dynamic time is required for disk fragmentation \citep[for a review, see][]{durisen07}. The questions are whether cooling by radiative transport with realistic opacities is rapid enough anywhere in protoplanetary disks to cause fragmentation and whether clumps that do form remain physically intact, gravitationally bound entities before the gas disk dissipates \citep[e.g.,][]{haisch01}.  
There have been sharp disagreements on these points among the different groups attacking the problem, and so it is important to evaluate how and why different methodologies may lead to strikingly different conclusions.

For moderate mass disks with a radial extent of tens of AU around solar-type stars, simulations with radiative transport and realistic opacities presented in \citet[][hereafter B07]{boss07} support disk fragmentation, in partial agreement with \cite{mayer07} but in  severe disagreement with simulations by our own group \citep{boley06, boley07, cai06, cai08, bd08} and others \citep{stam08,forg09}. In B07, Boss considers a variety of factors that may  account for the disagreement between his results and ours.
We as well have explored some of the issues described in B07, including:  resolution  \citep{pickett03, bd08}, radiative transport algorithms \citep{boley07}, irradiation \citep{cai08}, and viscosity \citep{pickett07}.  In this paper, we focus on  differences in the radiative schemes by using initial conditions and input physics that are as similar as possible to those described in B07 and by keeping as much of the numerical treatment as close to the techniques and conditions regularly used by Boss.  By employing a different and, we think, better treatment of the radiation physics, we find, as we have before, that realistic cooling is not nearly strong enough to initiate clump formation.

This paper is organized as follows.  In Section 2, we describe our numerical methodology and initial axisymmetric equilibrium state.  We describe the results for a simulation lasting five outer rotation periods in Section 3.  In Section 4, we compare our results with those in B07 and try to isolate the causes for our differences.
Section 5 is a brief summary.

\section{Computational Methodology}

\subsection{3-D Radiative Hydrodynamics}

We conducted our three-dimensional disk simulation using the CHYMERA (Computational HYdrodynamics with MultiplE Radiation Algorithms) code, 
developed at Indiana University  \citep[e.g.,][]{boley07}. CHYMERA uses  an Eulerian scheme that is fully second-order in space and time to solve the equations of hydrodynamics and  Poisson's equation on a cylindrical grid.  The grid has a resolution of (256,512,64) in cylindrical coordinates ($r$, $\phi$, $z$).  The disk is initially located between radial cells 40 and 202, corresponding to 4 and 20 AU, but is free to expand both radially and vertically.  Hydrodynamic boundary conditions are outflow along the top, sides and inner hole of the grid;  material from the disk that moves inside the inner radius at radial cell number 13 is automatically added to the central star's mass.  Equatorial plane symmetry is assumed. 

The primary difference between the version of CHYMERA used in this paper and the one used in \citet{boley06} and \citet{cai06,cai08} is an improved treatment of the radiative physics in the optically
thin region \citep[for more details and test results, see][]{zhu09}; Cai et al. 2010, in preparation). 

Define $\tau$ as the optical depth measured vertically downward from above, calculated for each computational cell using Rosseland mean opacities. Regions in which $\tau \geq$ 2/3 are part of the disk interior, and we calcuate the radiant energy flow in all three directions using flux-limited diffusion with Rosseland mean opacities.  
On the other hand, we  allow the atmospheric cells (in which $\tau <$  2/3) to cool radiatively via an optically thin LTE emissivity  based on the local temperature but computed relative to radiative equilibrium. 
Thus, the  volumetric cooling rate in any atmospheric cell  is
\begin{equation}
\Lambda = 4\rho \kappa(T)(\sigma T^4 - \pi J), 
\end{equation}
where $\kappa(T)$ is the Planck mean opacity and $J$ is the mean intensity, given by
\begin{equation}
\pi J = \frac{3\sigma}{4}T_{ph}^4(\tau+2/3)+\sigma T_{env}^4, 
\end{equation}
which is a radiative equilibrium solution for an Eddington-like grey atmosphere including 
external irradiation with a downward flux at the upper boundary of $\sigma T^4_{env}$. We couple the optically thick and thin regions using  a boundary condition, which defines  the $\tau = 2/3$ boundary flux $\sigma T_{ph}^4$ at the photosphere for the interior diffusion approximation. 

This scheme effectively measures optically thin local heating and cooling relative to the radiative equilibrium represented by $J$ and avoids the discontinuity at the photosphere and the artificially reduced atmospheric temperatures reported in \cite{boley06} for an earlier and cruder scheme. 
A similar radiative scheme was successfully employed to study FU Orionis outbursts \citep{zhu09}.

In B07, on the other hand, no radiative physics is done in the optically thin regions. Instead, the temperature above the photosphere is reset to that of the background temperature of an assumed thermal bath every time step, without computation of the detailed heating and cooling of optically thin cells. The boundary temperature for the radiative diffusion solution in the disk's optically thick interior is then taken to be the background temperature.

The opacities and molecular weights in our simulations are described in \cite{dalessio01} and are similar to those used in the Boss calculations.  
In the present case,  we adopt a revised mean molecular weight of $\mu$ = 2.39  \citep[lowered from the value in][]{cai08} for the temperature and pressure ranges in our disk simulation, close to  the $\mu$-value used  in B07 \citep{boss01}.  As in \cite{cai08}, we set the maximum particle radius in the dust opacity to one micron, compatible with B07.   
Note for clarity that we are {\sl not} using the option in CHYMERA where the entire $z$-direction is computed using radiative transfer along a single ray \citep[e.g.,][]{boley07, bd08}.

Equation (2) includes irradiation as a downward black-body flux  at $\tau = 0$ from an envelope with temperature $T_{env}$ = 50 K.   We do this to mimic the {\sl thermal bath} approach in B07, where the optically thin regions are maintained at a fixed temperature, typically 50 K. We point out that the effect of the thermal bath in B07 is to set a temperature boundary condition (BC) for the diffusive optically thick region, rather than a radiative flux BC to couple the thick and thin layers, as we do. As far as we know, the photospheric BC and the explicit radiative treatment of the $\tau <$ 2/3 region in our code represent the only substantive differences between the physics included in the Boss simulations and our own.  One minor difference in the interior treatment is that we have enabled the flux limiter, which, according to \cite{boss01}, should not change the outcome. On the numerical side, the B07 grid is spherical, not cylindrical, so the disk photosphere in B07 is located by a $\tau$ condition along spherical radii not the $z$-direction.

\subsection{Initial Model and Simulation Conditions}

The protoplanetary disk system is comprised of a star with one solar mass and a disk of material in nearly Keplerian rotation
with a mass equal to 0.091 $M_\odot$.  The initial model uses the \cite{boss84} pressure equation of state (EOS) based on an initial temperature profile provided by Boss (2004, private communication), but we derive the specific internal energy density by using $ \epsilon = p/ (\gamma - 1)$, where $\gamma$ is the ratio of specific heats and $\epsilon$ is the internal energy density. 
In the subsequent evolution, the EOS is assumed to be that of an ideal gas with $\gamma = 5/3$.
As in previous simulations, the disk begins in a marginally unstable state and is allowed to cool radiatively to instability. The initial axisymmetric model is set up using the same analytic expressions given in \cite{boss03}, but without infall. The latter is not a serious omission, because, in the B07 simulation, there is little mass in the inflow, and infall lasts only $\sim$ free-fall time, about 0.2 ORP. We included this short accretion phase in preliminary simulations with the \cite{cai08} version of CHYMERA and obtained essentially the same results reported here.
We introduce the same initial perturbation used in B07 which preferentially seeds power at the percent level into ordered ${\rm cos}(m\phi)$ structure with $m$ = 2 to 4 plus a small random component \citep{boss98,boss00}.  We ran the simulation 250,000 computation steps, equal to 5.0 ORPs.  An ORP (outer rotation period)
is the orbital period for material initially located at 20 AU, or about 90 yrs. The end of the simulation extends a little more than one ORP beyond the time shown in Figures 2 and 3 in B07.   Though material is allowed to expand off the grid (and  accrete onto the central star), the disk itself loses less than 0.4 $\%$ of its original mass and angular momentum. The simulation was conducted on a dedicated HP Proliant DL385 G2 server at Lawrence University.

\section{Results}

Figure 1 shows a three panel equatorial plane mass density contour map at three different times in the simulation.  In each case, the density contours span about 7 orders of magnitude, with each contour level representing a factor of 2 change in density; filled contours indicate regions of the highest density ($\rho \geq$ 10$^{-10}$ gcm$^{-3}$, following the convention  of B07).  The box for each image has dimensions of 40 AU $\times$ 40 AU.  The top panel shows the model at the time  nonaxisymmetric structure reaches its maximum strength, around t = 0.25 ORP.   Note the strong low-order spiral disturbance (primarily four-armed) near the inner boundary,  which is a result of the initial perturbation. In the middle panel, after 3.7 ORP's, only a remnant of nonaxisymmetric structure is seen in the bulk of the disk. The highest density regions and largest amplitude disturbances 
are both located near the inner boundary of the grid, though some spiral features can be traced lightly in the outer regions. This is completely different from what is shown at the same evolutionary time for models H and TZ in Figures 2 and 3 of B07, namely, dense clumps and/or nascent clumping near 10 AU. Over the final two ORP's of our simulation, as shown in the bottom panel of Figure 1, our disk has evidently settled into a quasi-steady configuration, consisting primarily of a  dense ring of material near the inner hole and 
a large but low-amplitude
two-armed spiral, parts of which extend beyond the original 20 AU edge of the disk.  Overall, the primary effect of the GIs is radial transport of mass.

Unlike some of our previous simulations, in which an unstable disk fragments into short-lived clumps \citep{pickett03, mejia05, durisen08}, no clumps ever form during the course of  this simulation.  Instead, the initial 2-, 3- and 4-armed perturbations grow briefly and then decay.  Although the nonaxisymmetric structure reaches a modest nonlinear  level, with amplitudes in $\delta \rho/ \rho \sim$ 0.1 for $m$ = 2, 3 and 4 at about 0.25 ORP,
no clumps form.  
Local cooling times $t_{\rm cool}$ for vertical columns through the disk in units of the local disk orbital period $P_{\rm rot}$ vary from about 15 to 40 across the most active regions of the disk (radial distances of 10 AU or less). The \cite{gammie01} fragmentation criterion, as applied to a $\gamma = 5/3$ EOS in a 3D simulation, leads us to expect that $t_{\rm cool}/P_{\rm rot}$ must be less than one for these regions to fragment \citep{rice03,rice05}. Seeing no clump formation is consistent with our high values of $t_{\rm cool}/P_{\rm rot}$.

We also investigated whether our disk is convectively unstable, because convection has been proposed \citep{boss04, mayer07} as an efficient cooling mechanism responsible for fragmentation.
Our Schwarzschild criterion analysis shows that only small regions of the disk are susceptible to convective instability early in the evolution.  By 1 ORP,  these regions are confined to thin strips not associated with the dynamics of material near the equatorial plane.

\section{Discussion}

Our simulation  results are clearly different from those highlighted  in B07.  All four simulations described  in B07 have the same grid resolution (512 azimuthal cells) as ours. The B07 models 
H and TZ also use the  same 50 K thermal bath BC.  While other parameters are varied in B07 (for example, the condition of a free vertical temperature gradient or one forced to be monotonically decreasing), the variation in thermal treatment of disk models in B07 does not greatly affect the ultimate outcome of clump formation.

As Boss notes in B07 and further explores, there are a number of possible reasons for the differences between his simulations and  ours. 
Here, we have attempted to remove most of them. For example, because B07 did not implement an artificial viscosity, we turned off CHYMERA's 
artificial viscosity for the simulation reported here. We used the same disk, the same EOS, and the same initial perturbation as in B07. 
We also include envelope irradiation to mimic the B07 thermal bath in a physically realistic manner. 
Although our code uses fewer terms in the boundary potential expansion than in B07, studies discussed in several of our  papers \citep{pickett03, pickett07, boley07, bd08} show that our boundary potential expansion is adequate and that the 512 azimuthal cells used here are sufficient to detect fragmentation when it should occur \citep[see, esp.,][]{bd08}.  
Moreover, the numerical heating in the interior disk regions that affected some of our previous simulations \citep{cai06, boley06} has been eliminated by the increased effective vertical resolution of the current model, in which we typically resolve  the vertical disk structure with  at least 10 cells inside 10 AU and with many more near and outside 10 AU. Tests in \cite{boley07} indicate that the disk must be vertically resolved by at least about 5 or 6 cells above the midplane to avoid numerical difficulties, including possible artificial heating.

Given the similarity of the models and input physics and given the elimination of all other major numerical concerns, the radical disagreement in results between our simulation and those in B07 must be due to differences in treatment of the photospheric BCs and the optically thin layers above the photosphere. 
\citet{boss04} recognizes that rapid
cooling is needed for fragmentation and attributes this to thermal convection, which he and also \citet{mayer07} identify with upwellings at close to the sound speed associated with the spiral arms. The upwellings are actually hydraulic jumps due to disruption of vertical hydrostatic equilibrium in spiral shocks \citep{bd06} and should not produce rapid cooling when properly modeled \citep{rafikov07,boley06,boley07,bd08}. In B07, however, the temperature excess over the thermal bath of any shock-heated material moving upward across the disk photosphere is effectively set to zero. To test how this affects cooling, we measured the flux of excess thermal energy in our post-shock regions at the time when we see maximum spiral wave amplitudes (top panel of Figure 1). We estimate that, in the densest shock structures, removal of this excess thermal energy as the gas jumps above the photosphere would produce an effective local $t_{\rm cool}/P_{\rm rot}$ of 2 to 3, compared with the true radiative $t_{\rm cool}/P_{\rm rot}$ of more like 15 at the same places, a reduction in local cooling time by a factor of between 5 and 8! While $t_{\rm cool}/P_{\rm rot}$ is not yet small enough to induce fragmentation, the B07 simulations do evolve denser structure and stronger shocks than our simulation beyond this time, and it is easy to imagine that the Gammie fragmentation criterion could then be satisfied. This estimate for the excess cooling introduced by the B07 temperature reset in the optically thin regions is highly suggestive that fast cooling in the B07 scheme is artificial. Unfortunately, when we tried to implement the B07 BCs in our own code, we encountered numerical instability in the radiative heating and cooling terms near the photosphere, and we could not test this hypothesis any further.

So what then is the status of disk instability as a gas giant formation mechanism?
Although \cite{mayer07} agree with Boss that disk instability may work under some conditions inside 40 AU, simulations by other groups \citep{stam08,forg09} and analytic treatments \citep{rafikov05,rafikov07} support our arguments against fragmentation in the inner disks of young solar-type stars. On the other hand, recent hydrodynamics simulations \citep{stam09,boley09,boleyetal09,vb10,hayfield10} and other arguments \citep{clarke09,rafikov09,dodson09} now suggest that disk instability may very well operate in the outer regions of large ($>$ 100 AU) massive disks during the early accretion phase to produce super-Jupiters, brown dwarfs, and low-mass stars. We think these fragmentation results in outer massive disks are quite plausible and worthy of further study.
The disk instability mechanism that Boss has championed since 1997 may yet prove to be an important formation channel for gas giants, but in a different place than he originally proposed.

\section{Summary}

We have conducted a high resolution simulation of a 4 to 20 AU protoplanetary disk of 0.091 $M_{\odot}$ orbiting a 1 $M_{\odot}$ star similar in most respects to simulations highlighted in B07.  We have eliminated as many differences in methodology as  possible, and we have answered the main arguments  in B07 against the robustness of our 
previous results.  The simulations in B07 show fragmentation into dense clumps; the simulation we present here shows no fragmentation.  
Because everything else is the same, the disagreement must be due to differences in the 
treatment of the optically thin regions and the photospheric boundary conditions. We identify at least one possible source of artificially fast cooling in the B07 scheme and demonstrate that it can decrease cooling times by a large factor. This work supports the conclusion that gas giant planet formation will not occur via disk instability in the inner tens of AU of disks around solar-type stars.

\section{Acknowledgments}

We thank A. C. Boley, D. Burrows, L. Mayer, S. Michael, and T.Y. Steiman-Cameron for useful scientific discussions and support. Special thanks are due to A.P. Boss who patiently worked with us to reproduce his initial conditions and input physics accurately. We also thank the anonymous referees whose comments led to substantial improvements of the manuscript. This research was conducted with the support of NASA grants from the Origins of Solar Systems  Program (NNG05GN11G and NNX08AK36G) and the Planetary Geology and Geophysics Program (NNX08AU41G) .

\newpage

\begin{center}
{\large\bf Figure Captions}
\end{center}

{\bf Figure 1.}  Equatorial plane mass density contours for our disk simulation, in a format similar to B07.  The contours span approximately 7 orders of magnitude in mass density, using factor of two spacing.  The box that encloses each image spans 40 AU on each side, accommodating the initial disk radius of 20 AU.  Shown are density images near the time of the highest amplitude nonaxisymmetry (top, 0.25 ORP), the  time at the end of the equivalent  runs in B07 (middle, 3.7 ORP), and the final image (bottom, 5.0 ORP). As in B07, we indicate the highest densities ($>$ 10$^{-10}$ gcm$^{-3}$) with filled black contours.

\label{lastpage}


\begin{thebibliography}{99}

\bibitem[\protect\citeauthoryear{Boley}{2009}]{boley09} 
Boley,  A.C. 2009, ApJ, 695, L53 

\bibitem[\protect\citeauthoryear{Boley \& Durisen}{2006}]{bd06} Boley, A.C., Durisen, R.H. 2006, ApJ, 641, 534

\bibitem[\protect\citeauthoryear{Boley \& Durisen}{2008}]{bd08} Boley, A.C., Durisen, R.H. 2008, ApJ, 685, 1193

\bibitem[\protect\citeauthoryear{Boley et al.}{2006}]{boley06} Boley, A.C., 
Mej\'ia, A.C., Durisen, R.H., Cai, K., Pickett, M.K. 2006, ApJ, 651, 517

\bibitem[\protect\citeauthoryear{Boley et al.}{2007}]{boley07} Boley, A.C., 
Durisen, R.H., Nordlund, A., Lord, J. 2007, ApJ, 665, 1254 

\bibitem[\protect\citeauthoryear{Boley et al.}{2009}]{boleyetal09} Boley, A.C., Hayfield, T., Mayer, L., Durisen, R.H.  2010, Icarus, 207, 509 (arXiv:0909.4543)

\bibitem[\protect\citeauthoryear{Boss}{1984}]{boss84} Boss, A.P. 1984, ApJ, 277, 768

\bibitem[\protect\citeauthoryear{Boss}{1997}]{boss97} Boss, A.P. 1997, Science, 267, 1836

\bibitem[\protect\citeauthoryear{Boss}{1998}]{boss98} Boss, A.P. 1998, ApJ, 503, 923

\bibitem[\protect\citeauthoryear{Boss}{2000}]{boss00} Boss, A.P. 2000, ApJ, 536, L101

\bibitem[\protect\citeauthoryear{Boss}{2001}]{boss01} Boss, A.P. 2001, ApJ, 563, 367

\bibitem[\protect\citeauthoryear{Boss}{2002}]{boss02} Boss, A.P. 2002, ApJ, 576, 462

\bibitem[\protect\citeauthoryear{Boss}{2003}]{boss03} Boss, A.P. 2003, ApJ, 599, 577

\bibitem[\protect\citeauthoryear{Boss}{2004}]{boss04} Boss, A.P. 2004, ApJ, 610, 456

\bibitem[\protect\citeauthoryear{Boss}{2005}]{boss05} Boss, A.P. 2005, ApJ, 629, 535

\bibitem[\protect\citeauthoryear{Boss}{2007}]{boss07} Boss, A.P. 2007, ApJ, 661, L73

\bibitem[\protect\citeauthoryear{Boss}{2009}]{boss09} Boss, A.P. 2009, ApJ, 694, 107

\bibitem[\protect\citeauthoryear{Cai et al.}{2006}]{cai06} Cai, K., Durisen, R.H., Michael, S., Boley, A.C., Mej\'ia, A.C., Pickett, M.K., D'Alessio, P. 2006, ApJ, L149

\bibitem[\protect\citeauthoryear{Cai et al.}{2008}]{cai08} Cai, K., Durisen, R.H., Boley, A.C., Pickett, M.K., Mej\'ia, A.C. 2008, ApJ, 673, 1138

\bibitem[\protect\citeauthoryear{Clarke}{2009}]{clarke09} 
Clarke, C.J. 2009, MNRAS, 396, 1066

\bibitem[\protect\citeauthoryear{D'Alessio et al.}{2001}]{dalessio01} D'Alessio, P., Calvet, N., Hartmann, L. 2001, ApJ, 553, 321

\bibitem[\protect\citeauthoryear{Dodson-Robinson et al.}{2009}]{dodson09}
Dodson-Robinson, S. E., Veras, D., Ford, E. B., Beichman, C. A. 2009, ApJ, 707, 79

%\bibitem[\protect\citeauthoryear{Durisen, Mej\'ia, \& Pickett}
%{Durisen et al.}{2003}]{durisen03} Durisen, R.H., Mej\'ia, A.C., 

\bibitem[\protect\citeauthoryear{Durisen et al.}{2007}]{durisen07} Durisen R.H., Boss, A.P., Mayer, L., Nelson, A.F., Quinn, T., Rice, W.K.M. 2007, in: Protostars and Planets V, ed. B. Reipurth, D. Jewitt, \& K. Keil (Tucson:  Univ. of Arizona Press), 607

\bibitem[\protect\citeauthoryear{Durisen et al.}{2008}]{durisen08} 
Durisen R.H., Hartquis, T.W., \& Pickett M.K., 2008, Ap \& SpSci., 317, 3 

\bibitem[\protect\citeauthoryear{Forgan et al.}{2009}]{forg09} Forgan, D., Rice, K., Stamatellos, D., Whitworth, A. 2009, MNRAS 394, 882

\bibitem[\protect\citeauthoryear{Gammie}{2001}]{gammie01} Gammie, C.F. 2001, ApJ, 553, 174 

\bibitem[\protect\citeauthoryear{Johnson \& Gammie}{2003}]{johnson03} 
Johnson, B.M., Gammie, C.F. 2003, ApJ, 597, 131 

\bibitem[\protect\citeauthoryear{Haisch et al.}{2001}]{haisch01} 
Haisch, K.E., Jr., Lada, E.A., Lada, C.J. 2001, ApJ, 553, L153

\bibitem[\protect\citeauthoryear{Hayfield et al.}{2010}]{hayfield10}
Hayfield, T., Mayer, L., Wadsley, J., Boley, A. C. 2010, MNRAS, submitted (arXiv:1003.2594)

\bibitem[\protect\citeauthoryear{Mayer et al.}{2002}]{mayer02} Mayer, L., Quinn, T. Wadsley, J., Stadel, J. 2002, Science, 298, 1756

\bibitem[\protect\citeauthoryear{Mayer et al.}{2004}]{mayer04} Mayer, L., Quinn, T., Wadsley, J., Stadel, J. 2004, ApJ, 609, 1045 

\bibitem[\protect\citeauthoryear{Mayer et al.}{2007}]{mayer07} 
Mayer, L., Lufkin, G., Quinn, T., Wadsley, J. 2007, ApJ, L77 

\bibitem[\protect\citeauthoryear{Mej\'ia et al.}{2005}]{mejia05} Mej\'ia, A.C., Durisen, R.H., Pickett, M.K., Cai, K. 2005, ApJ, 619, 1098 

\bibitem[\protect\citeauthoryear{Nelson}{2006}]{nelson06} Nelson A.F. 2006, MNRAS, 373, 1039

\bibitem[\protect\citeauthoryear{Nelson et al.}{1998}]{nelson98} Nelson, A.F., Benz, W., Adams, F.C., Arnett, D. 1998, ApJ, 502, 342 

\bibitem[\protect\citeauthoryear{Nelson, Benz, Ruzmaikina}{Nelson et al.}{2000}]
{nelson00} Nelson, A.F., Benz, W., Ruzmaikina, T.V. 2000, ApJ, 529, 357 

\bibitem[\protect\citeauthoryear{Pickett et al.}{1998}]{pickett98} Pickett, B.K., Cassen, P., Durisen, R.H., Link, R. 1998, ApJ, 504, 468

\bibitem[\protect\citeauthoryear{Pickett et al.}{2000a}]{pickett00a} Pickett, B.K., Cassen, P., Durisen, R.H., Link, R. 2000, ApJ, 529, 1034

\bibitem[\protect\citeauthoryear{Pickett et al.}{2000b}]{pickett00b} Pickett, B.K., Durisen, R.H., Cassen, P., Durisen, R.H., Mej\'ia, A.C. 2000, ApJ, 540, L95

\bibitem[\protect\citeauthoryear{Pickett et al.}{2003}]{pickett03} Pickett, B.K., Mej\'ia, A.C., Durisen, R.H., Cassen, P., Berry, D.K., Link, R.P. 2003, ApJ, 590, 1060

\bibitem[\protect\citeauthoryear{Pickett \& Durisen}{2007}]{pickett07} 
Pickett, M.K., Durisen, R.H. 2007, ApJ, 654, L155 

\bibitem[\protect\citeauthoryear{Rafikov}{2005}]{rafikov05} 
Rafikov,  R.R. 2005, ApJ, 621, L69 

\bibitem[\protect\citeauthoryear{Rafikov}{2007}]{rafikov07} 
Rafikov,  R.R. 2007, ApJ, 662, 642 

\bibitem[\protect\citeauthoryear{Rafikov}{2009}]{rafikov09} 
Rafikov,  R.R. 2009, ApJ, 704, 281

\bibitem[\protect\citeauthoryear{Rice et al.}{2003}]{rice03} Rice, W.K.M., Armitage, P.J., Bate, M.R., Bonnell, I.A. 2003, MNRAS, 339, 1025

\bibitem[\protect\citeauthoryear{Rice et al.}{2005}]{rice05} Rice, W.K.M., Lodato, G., Armitage, P.J. 2005, MNRAS, 364, L56 

\bibitem[\protect\citeauthoryear{Stamatellos \& Whitworth}{2008}]{stam08} Stamatellous, D., \& Whitworth, A. 2008, A\& A, 480, 879

\bibitem[\protect\citeauthoryear{Stamatellos \& Whitworth}{2009}]{stam09} Stamatellous, D., \& Whitworth, A. 2009, MNRAS, 392, 413

\bibitem[\protect\citeauthoryear{Vorobyov \& Basu}{2010}]{vb10}
Vorobyov, E.I., Basu, S. 2010, ApJ, 714, L133

\bibitem[\protect\citeauthoryear{Zhu et al.}{2009}]{zhu09} Zhu, Z., Hartmann, L., Gammie, C.F.; McKinney, J.C. 2009, ApJ, 701, 620


\end{thebibliography}
\end{document}